\documentclass[12pt]{iopart}
\usepackage{iopams}
\usepackage{graphicx}
\usepackage{epsf}
\usepackage{dcolumn}
\usepackage{bm}
\begin{document}

\title{Thermal emission in the ultrastrong coupling regime}

\author{A. Ridolfo$^1$, M. Leib$^1$, S. Savasta$^2$, and M. J. Hartmann$^1$}

\address{$^1$Physik Department, Technische Universit\"{a}t M\"{u}nchen, James-Franck-Strasse, 85748 Garching, Germany\\
             $^2$Dipartimento di Fisica della Materia e Ingegneria Elettronica, Universit\`{a} di Messina Salita Sperone 31, I-98166 Messina, Italy}
\ead{alessandro.ridolfo@ph.tum.de}

\begin{abstract}
We study thermal emission of a cavity quantum electrodynamic system in the ultrastrong-coupling regime where the atom-cavity coupling rate becomes comparable the cavity resonance frequency.
In this regime, the standard descriptions of photodetection and dissipation fail. Following an approach that was recently put forward by Ridolfo et al.[arXiv:1206.0944], we are able to calculate the emission of systems with arbitrary strength of light matter interaction, by expressing the electric field operator in the cavity-emitter dressed basis. Here we present thermal photoluminescence spectra, calculated for given temperatures and for different couplings in particular for available circuit QED parameters.
\end{abstract}

\maketitle

\section{Introduction}
The quantum theory of photodetection as originally formulated by Glauber \cite{Glauber} is a landmark for quantum optics and has occupied a key role in understanding radiation-matter interactions. Recently, a new regime of interaction cavity quantum electrodynamics (cavity QED) has been reached experimentally where the coupling strength between an emitter and the cavity photons becomes comparable to the transition frequency of the emitter or the resonance frequency of the cavity mode \cite{guenter09,Niemczyk,Todorov,Schwartz,Hoffman, Scalari}. In this so called ultrastrong coupling (USC) regime \cite{De Liberato,Peropadre2010,Nataf,guenter09,Niemczyk,Casanova} the usual rotating wave approximation is no longer applicable. As a consequence the number of excitations in the cavity-emitter system is no longer conserved, even in the absence of drives and dissipation.
Hence in this regime, the standard descriptions of photoluminescence of a strongly coupled system fails \cite{Portolan,Ridolfooecs,RidolfoPRB}.
According to the very recent result \cite{Ridolfo}, it is possible to calculate the correct output radiation avoiding unphysical contributions to the emission from the vacuum due to the presence of the counter rotating terms. Here we present the output thermal emission for a cavity QED system where a two level system (TLS) interacts with one cavity mode in the ultrastrong coupling regime. In particular, we analyze the case in wich the interaction has the form of the Rabi Hamiltonian, showing how the spectra modify their shape as the coupling strength or the thermal feeding increases.
Our results can be measured in a wide recent experimental setups \cite{guenter09,Niemczyk,Todorov,Schwartz,Hoffman, Scalari}.
\section{Input-output relations}
Applying Glauber's idea of photodetection, we here introduce a full quantum theory to study the thermal emission in the USC regime \cite{Ridolfo}.
This requires a proper generalization of input-output theory  \cite{GardinerZoller}, since the standard relations would for example predict an output photon flux that is proportional to the average number of cavity photons, i.e. $\langle a_{\rm out}^\dag(t) a_{\rm out}(t) \rangle \propto \langle a^\dag(t) a(t) \rangle$ for vacuum input. Hence an unwary application of this standard procedure to the USC regime would predict an unphysical stream of output photons for a system in its ground state which contains a finite number of photons due to the counter-rotating terms in the Hamiltonian.
It has been shown \cite{Ridolfo} that applying the Glauber's formulation of photodetection \cite{Glauber}, it is possible to derive n-\emph{th} order correlation functions for the output fields which are valid for arbitrary degrees of light-matter interaction, by expressing the cavity electric-field operator in the atom-cavity dressed basis.
An ideal detector absorbs a photon with a probability per unit time that is proportional to $\langle E^-(t) E^+(t) \rangle$ where $E^\pm(t)$ are the positive and negative frequency components of the electric field operator of the output field \cite{Glauber,Milonni}. In the circuit QED the same quantities are measured with output voltages which are proportional to the electric fields.
Following \cite{Ridolfo}, the input-output relations for a cavity that is coupled to a one-dimensional output waveguide via an interaction between the cavity field $X$ and the momentum quadratures $\Pi_{\omega}$ of the waveguide field outside the cavity are
\begin{equation} \label{eq:input-output}
a_{\rm{out}}(t) = a_{\rm{in}}(t) - i \frac{\epsilon_{c}}{\sqrt{8 \pi^{2} \hbar \epsilon_{o} v}} \dot{X}^{+},
\end{equation}
where $\epsilon_{c}$ is a coupling parameter and $\epsilon_{o}$ is a parameter describing the dielectric properties of the output waveguide, $v$ is the phase velocity.
The input(output) field operators $a_{\rm{out}(\rm{in})}(t)$ are defined as
\begin{equation} \label{eq:output-def}
a_{\rm{out}(\rm{in})}(t) = \frac{1}{\sqrt{2 \pi}} \int d\omega \sqrt{\omega} e^{-i \omega (t - \tilde{t})} a_{\omega}(\tilde{t}) ,
\end{equation}
where $\tilde{t} = t_{1} \to  +\infty$ for the output field and $\tilde{t} = t_{0} \to  -\infty$ for the input field and $a_{\omega}$($a_{\omega}^{\dagger}$) annihilation(creation) operators of the fields outside the cavity.
In this way, $\hbar \langle a_{\rm{out}}^{\dagger}(t) a_{\rm{out}}(t) \rangle$ is proportional to the measured $\langle E^-(t) E^+(t) \rangle$ and describes an energy flux associated to the output light.
The standard definition of output fields as used in many textbooks, c.f. \cite{Walls}, is recovered if all frequencies of the field are very close to a carrier frequency $\overline{\omega}$ and one may approximate $\sqrt{\omega} \approx \sqrt{\overline{\omega}}$ in the integral kernel which makes the observed energy fluxes proportional to photon number fluxes.
One thus needs to find the positive frequency component of $\dot{X}$ according to its actual dynamical behavior, c.f. \cite{Savasta96}, to compute the proper output fields. We do this by expressing $X$ in the atom-cavity dressed basis. It is worth to notice that in the USC regime, the positive frequency component of $X$ is not proportional to the photon annihilation operator $a$.
\section{Dynamics of the open quantum system}
We consider the Rabi model, that consists of a linear coupling between a single cavity mode and a two level system (TLS). In the following we set for sake of simplicity $\hbar = 1$ and also the Boltzmann constant $k_{B} = 1$. The Rabi Hamiltonian reads
\begin{equation}\label{eq:model}
    H_{\rm S}  = \omega_{\rm 0} a^{\dagger}a + \omega_{\rm x} \sigma^{+} \sigma^{-} + g ( a + a^{\dagger})\sigma_{\rm x}
\end{equation}
where $\omega_{\rm 0(x)}$ is the bare energy of the cavity mode (TLS), $g$ is the coupling strength and $\sigma_{\rm x}$ is the standard Pauli operator.
\begin{figure}[!ht]
\centering
\includegraphics[height=50mm]{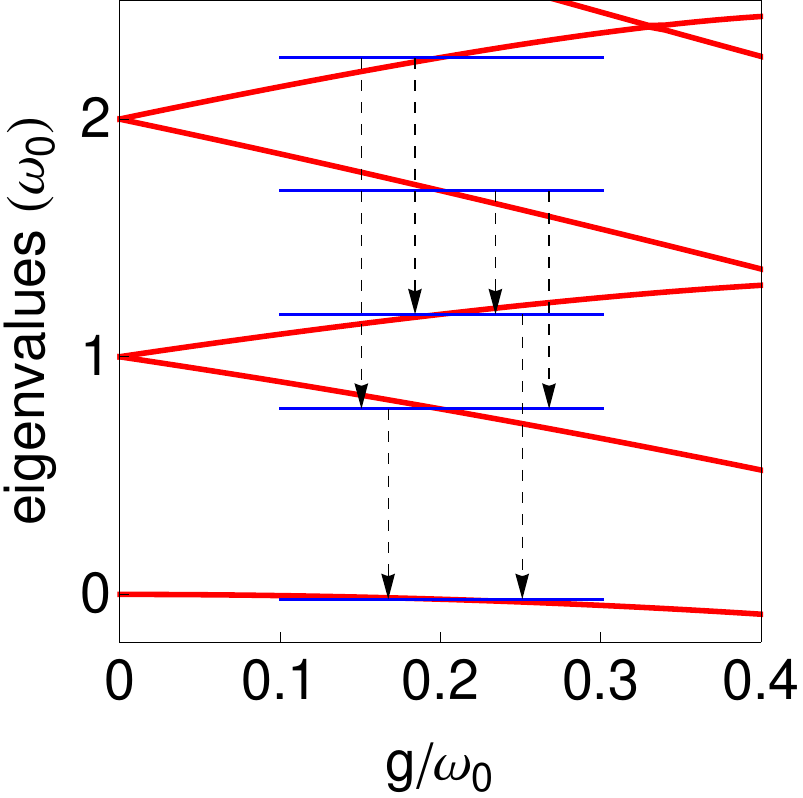}
\caption{(color online) Energy spectrum of $H_{\rm 0}$ as function of the coupling strength for the Rabi Hamiltonian. This plot is calculated for $\omega_{\rm x} = \omega_{\rm 0} $. The level structure is analogous to that of the JC model. The ladders correspond to a coupling strength $g = $ 0.2 $\omega_{\rm 0}$. The arrows indicate possible transitions of radiative decay due to the Hamiltonian structure.}
\label{fig:spectrum}
\end{figure}
Fig. \ref{fig:spectrum} shows a plot of the spectrum of the eigenvalues as function of the coupling $g$ calculated for the parameters $\omega_{\rm 0} = \omega_{\rm x} = 1$. In particular, it is worth to notice that for a given value of $g$, the system shows the characteristic ladder configuration like in the standard Jaynes-Cummings model. The possible transitions between the rungs are given by the selection rules due to the Hamiltonian structure, that enables these transitions owing to the non-zero values of the elctric field components. In order to describe a realistic system, dissipation induced by its coupling to the environment needs to be considered. Then, owing to the very high ratio $g/\omega_{\rm 0}$, a standard quantum optical master equations fails as it would for example predict that even zero temperature environments could drive the system out of its ground state. An adequate description of the system's coupling to its environment requires a perturbative expansion in the system-bath coupling 
strength.
In order to perform this expansion we write the Hamiltonian in a basis formed by eigenstates $|j\rangle$ of $H_{0}$, denote the respective energy eigenvalues by $\hbar \omega_j$, i.e. $H_{0}|j\rangle = \hbar \omega_j |j\rangle$, and derive Redfield equations \cite{breuer} to describe the dissipative processes \cite{Blais}. We choose the labeling of the states $|j\rangle$ such that $\omega_k > \omega_j$ for $k > j$ and focus on a single-mode cavity with $T\neq 0$ temperature for the environment. Generalizations to a multi-mode cavity are straightforward. For our porpouse, we neglect small Lamb shifts and dephasing contributions as they do not alter significantly the output photon, allowing a simpler and lighter theoretical setup. We thus arrive at the master equation,
\begin{equation}\label{master-eq}
    \dot\rho(t) = i [\rho(t), H_{\rm S}] + \mathcal{L}_{a}\rho(t) + \mathcal{L}_{x}\rho(t).
\end{equation}
The expressions $\mathcal{L}_{a}$ and $\mathcal{L}_{x}$ are Liouvillian superoperators describing the losses of the system where $\mathcal{L}_{c}\rho(t) = \sum_{j,k>j}\Gamma^{j k}_{c} (1+\bar{n}_{c}(\Delta_{k j},T)) \mathcal{D}[|j \rangle \langle k|]\rho(t) + \sum_{j,k>j}\Gamma^{j k}_{c} \bar{n}_{c}(\Delta_{k j},T) \mathcal{D}[|k \rangle \langle j|]\rho(t)$ for $c = a, \sigma^{-}$ and $\mathcal{D}[\mathcal{O}]\rho = \frac{1}{2} (2 \mathcal{O}\rho\mathcal{O}^{\dagger}-\rho \mathcal{O}^{\dagger} \mathcal{O} - \mathcal{O}^{\dagger} \mathcal{O}\rho)$ and $T$ is the temperature of the thermal bath. Here $\bar{n}_{c}(\Delta_{k j},T)$ is the number of thermal photons that feed the system acting on all the possible $|k \rangle\rightarrow | j \rangle$ transitions. Standard dissipators are recovered in the limit $g \to 0$. The relaxation coefficients $\Gamma^{j k}_{c} = 2\pi d_{c}(\Delta_{k j}) \alpha^{2}_{c}(\Delta_{k j})| C^{c}_{j k}|^2$ depend on the spectral density of the baths $d_{c}(\Delta_{k j})$ and the 
system-bath coupling strength $\alpha_{c}(\Delta_{k j})$ at the respective transition frequency $\Delta_{k j} = \omega_{k} - \omega_{j}$ as well as on the transition coefficients $C_{j k} = -i \langle j |(c - c^{\dagger})| k \rangle$ ($c = a, \sigma^{-}$). These relaxation coefficients can be interpreted as the full width at half maximum of each $|k \rangle\rightarrow | j \rangle$ transition. Since we consider a cavity that couples to the momentum quadratures of fields in one-dimensional output waveguides, it is possible to show \cite{Ridolfo} that the relaxation coefficients reduce to $ \Gamma^{j k}_{c} = \gamma_{c} \,\frac{\Delta_{k j}}{\omega_{0}} \, |C^{c}_{j k}|^2$, where $\gamma_{c}$ are the standard damping rates of a weak coupling scenario.

\section{Results}
According to the input-output relation (\ref{eq:input-output}) the output spectrum of light is, for input fields in vacuum, equal to,
\begin{equation}\label{spectrum}
S(\omega) \propto \lim_{t \to \infty} 2 \Re \int_{0}^{\infty} \langle \dot{X}^{-}(t)\dot{X}^{+}(t+\tau)\rangle e^{i \omega \tau} d\tau
\end{equation}
i.e. the Fourier transform of the two time correlation function. In Eq. (\ref{spectrum}) $X = -i X_{0} (a - a^{\dag})$ and $\Re$ denotes the real part.
In this way, we are calculating the photoluminescence of an interacting system that is thermalized at temperature $T$ and emits into the vacuum, i.e. into a reservoir at $T = 0$ temperature. To separate $\dot{X}$ in its positive and negative frequency components, $\dot{X}^{+}$ and $\dot{X}^{-}$, we expand it in terms of the energy eigenstates $|j\rangle$ and find $\dot{X}^+ = -i \sum_{j,k>j} \Delta_{kj}X_{jk} | j \rangle \langle k |$, where $X_{jk} = \langle j | X | k \rangle$ and $X^- = (X^+)^\dag$. Note that $X^+ |0 \rangle = 0$, for the system ground state $|0 \rangle$ in contrast to $a |0 \rangle \neq 0$. For a weak excitation density $S(\omega)$ is proportional to the emission. Then, once the master equation (\ref{master-eq}) is numerically solved, one can easily calculate the thermal spectrum (\ref{spectrum}) applying the quantum regression theorem \cite{Walls}.
\begin{figure}[!ht]
\centering
\includegraphics[height=50mm]{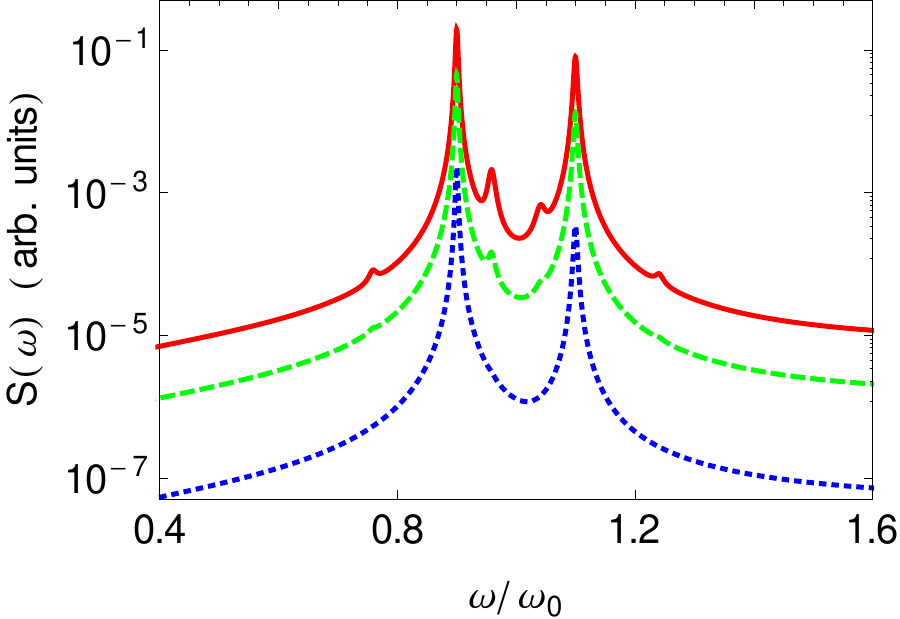}
\caption{(color online) Thermal emission spectra for different temperatures: $T = 0.1 \omega_{\rm 0}$ (blue-dotted line), $T = 0.15 \omega_{\rm 0}$ (green-dashed line), $T = 0.2 \omega_{\rm 0}$ (red-continuous line). The other parameters are $\omega_{\rm x} = \omega_{\rm 0}$, $\gamma_{a} = \gamma_{x} = 5 \times 10^{-3}$ $\omega_{\rm 0}$, and $g = 0.1$ $\omega_{\rm 0}$ for all three cases.}
\label{fig:sp01}
\end{figure}
Fig. \ref{fig:sp01} shows a plot of thermal emission spectra for different temperatures. The used parameters are scaled by the bare mode energy of the cavity and in this figure $g = 0.1$ $\omega_{\rm 0}$. If we choose $\omega_{\rm 0} = 2\pi \times 10$ GHz, the parameters belong to the common range of values of circuit QED systems and the corresponding temperatures fall in the region of hundreds of mK. In fact, for $\omega_{\rm 0} = 2\pi \times 10$ GHz, the respective temperature for the blue-dotted curve is 48 mK, for the green-dashed 72 mK, and for the red-continuous curve is 96 mK.
\begin{figure}[!ht]
\centering
\includegraphics[height=50mm]{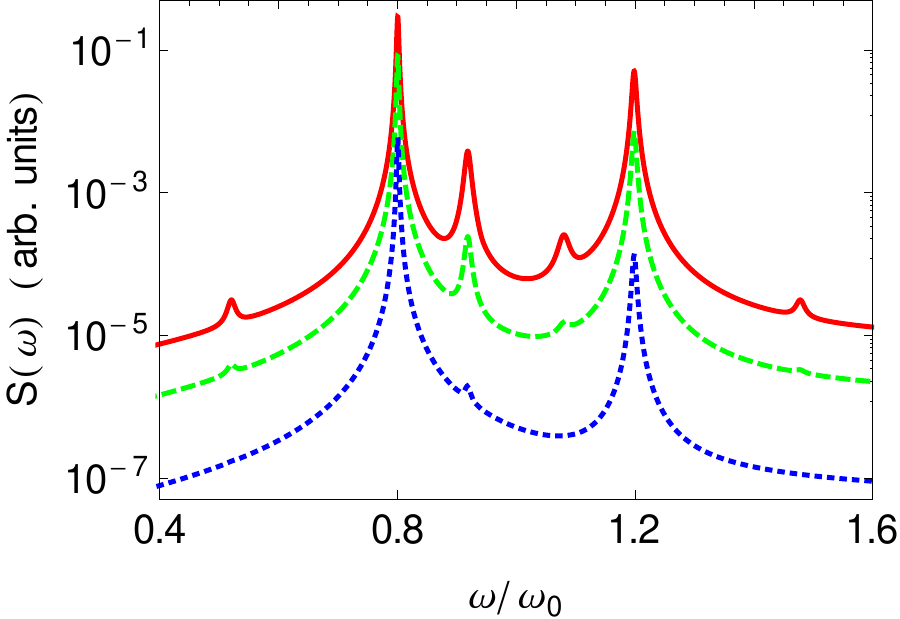}
\caption{(color online) Thermal emission spectra for different temperatures: $T = 0.1 \omega_{\rm 0}$ (blue-dotted line), $T = 0.15 \omega_{\rm 0}$ (green-dashed line), $T = 0.2 \omega_{\rm 0}$ (red-continuous line). The other parameters are the same used in Fig. \ref{fig:sp01}  except for the $g$ coupling that here is $g = 0.2$ $\omega_{\rm 0}$.}
\label{fig:sp02}
\end{figure}
In Fig. \ref{fig:sp02}, instead, we increased the value of the coupling up to $g = 0.2$ $\omega_{\rm 0}$, leaving the other parameters the same as in Fig. \ref{fig:sp01}. One can see that the mean peaks, that in Fig. \ref{fig:sp01} correspond to the values $\omega/\omega_{\rm 0} \sim 0.9$ and $\omega/\omega_{\rm 0} \sim 1.1$, are now separed roughly by twice the energy with respect to the previous case. This happens because the energy separation of the first two eigenmodes of (\ref{eq:model}), also in this range of interaction, growths linearly with respect to the coupling strength. Looking at the figures \ref{fig:sp01} and \ref{fig:sp02} we can see that if the temperature increases, the number of resonances present in the system increases as well. Indeed, whenever the temperature rises, this increases the thermal occupancy $\bar{n}_{c}(\Delta_{k j},T)$ allowing to exicite the $|k \rangle\rightarrow | j \rangle$ transitions. A characterization of the resonances, e.g. in Fig. \ref{fig:sp02}, is really 
straightforward if one looks at the ladder scheme (see Fig \ref{fig:spectrum}). Sorting the energies from the lower value to the higher, we identify $|4 \rangle\rightarrow | 3 \rangle$, $|2 \rangle\rightarrow | 1 \rangle$, $|4 \rangle\rightarrow | 2 \rangle$, $|5 \rangle\rightarrow | 3 \rangle$, $|3 \rangle\rightarrow | 1 \rangle$ and $|5 \rangle\rightarrow | 2 \rangle$ as shown by the rows in Fig. \ref{fig:spectrum}. It is worth to notice that in all the spectra that we presented, there is an evident asymmetry also in the absence of detunig, in contrast with the results achievable within the standard model for the dissipation and photodetection that naturally cannot apply in this regime of interaction. This is due to the fact that in the USC regime, i) the effect of the thermal feeding of the reservoir acts differetly on the different transitions and, ii) each resonance is characterized by a different damping rate.
\section{Conclusions}
In conclusion, we presented a full description of the thermal emission in the USC regime, valid for arbitrary light-matter couplings. We showed thermal emission spectra calculated for available circuit QED parameters. These results show that the recently proposed correlation function are also able to describe correctly incoherent light emission from USC systems avoiding unphysical emission from the ground state. Generalizations of
our study to multicavity
devices \cite{Leib10} and to three-level in the USC regime \cite{RidolfoPRL2011,odistefano1,StassiEPL2012} would form interesting perspectives
for future research.
\vspace{0.5 cm}



\begin{thebibliography}{99}


\bibitem{Glauber} R. J. Glauber, Phys. Rev. {\bf 130}, 2529 (1963).

\bibitem{guenter09} G. G\"unter {\em et al.}, Nature {\bf 458}, 178 (2009).

\bibitem{Niemczyk} T. Niemczyk {\em et al.}, Nat. Phys. {\bf 6}, 772 (2010).

\bibitem{Todorov} Y. Todorov {\em et al.}, Phys. Rev. Lett. {\bf 105}, 196402 (2010).

\bibitem{Schwartz} T. Schwartz, J.\ A. Hutchison, C. Genet, and T.\ W. Ebbesen, Phys. Rev. Lett. {\bf 106}, 196405 (2011).

\bibitem{Hoffman} A.J. Hoffman {\em et al.}, Phys. Rev. Lett. {\bf 107}, 053602 (2011).

\bibitem{Scalari} G. Scalari {\em et al.}, Science {\bf 16}, 1323 (2012).

\bibitem{De Liberato} S. De Liberato, C. Ciuti, and I. Carusotto, Phys. Rev. Lett. {\bf 98}, 103602 (2007).

\bibitem{Peropadre2010} B. Peropadre, P. Forn-Diaz, E. Solano, J.\ J.  Garc\'{i}a-Ripoll, Phys. Rev. Lett. {\bf 105}, 023601 (2010).

\bibitem{Nataf} P. Nataf, and C.  Ciuti, Phys. Rev. Lett. {\bf 104}, 023601 (2010).

\bibitem{Casanova} J. Casanova, G. Romero, I. Lizuain, J.\ J. Garc\'{i}a-Ripoll, and E. Solano, Phys. Rev. Lett. {\bf 105}, 263603 (2010).

\bibitem{Portolan} S. Portolan, O. Di Stefano, S. Savasta, F. Rossi, and R. Girlanda, Phys. Rev. B {\bf 77}, 035433 (2008).

\bibitem{Ridolfooecs} A. Ridolfo, O. Di Stefano, S. Portolan, and S. Savasta, J. Phys.: Conf. Ser. {\bf 210}, 012025 (2010).

\bibitem{RidolfoPRB} A. Ridolfo, S. Stelitano, S. Patan\`{e}, S. Savasta, and R. Girlanda, Phys. Rev. B {\bf 81},  075313 (2010).

\bibitem{Ridolfo} A. Ridolfo, M. Leib, S. Savasta, and M.J. Hartmann, arXiv:1206.0944v1 (2012).

\bibitem{GardinerZoller} C. W. Gardiner and P. Zoller, {\em Quantum Noise}, Springer-Verlag, (2000).

\bibitem{Milonni} P. W. Milonni and D. F. V. James, and H. Fearn, Phys. Rev. A {\bf 52}, 1525 (1995).

\bibitem{Walls} D. F. Walls and G. J. Milburn, Quantum Optics (Cambridge University Press, Cambridge, England, 1994).

\bibitem{Savasta96} S. Savasta and R. Girlanda, Phys. Rev. A {\bf 53}, 2716 (1996).


\bibitem{breuer} H.-P. Breuer and F.Petruccione, \emph{ The Theory of Open Quantum Systems}, Oxford University Press (2006).

\bibitem{Blais} F. Beaudoin, J. M. Gambetta, and A. Blais, Phys. Rev. A {\bf 84}, 043832 (2011).

\bibitem{Leib10} M. Leib and M.J. Hartmann, New J. Phys. {\bf 12}, 093031 (2010).

\bibitem{RidolfoPRL2011} A. Ridolfo, R. Vilardi, O. Di Stefano, S. Portolan, and S. Savasta, Phys. Rev. Lett. {\bf 106},
013601 (2011).

\bibitem{odistefano1} O. Di Stefano,  A. Ridolfo, S. Portolan, and S. Savasta, Opt. Lett. {\bf 36},  4509-4511 (2011).

\bibitem{StassiEPL2012}  R. Stassi, A. Ridolfo, S. Savasta, R. Girlanda, and O. Di Stefano, EPL {\bf 99},
24003 (2012).

\end{thebibliography}
\end{document}